\documentclass[journal]{IEEEtran} 

\usepackage[utf8]{inputenc}
 \usepackage{amsthm}
 \usepackage{mathtools}
\usepackage{amsfonts} 
\usepackage{amssymb} 
 \usepackage{setspace}
  \usepackage{mathrsfs}
  \usepackage{algorithm2e}
\usepackage{enumitem}
\usepackage[colorlinks=true,linkcolor=blue,citecolor=blue,urlcolor=blue]{hyperref}
\usepackage{hyperref}
 \usepackage{url}

\newcommand\sZ{\mathbb{Z}}
\newcommand\bR{\mathbb{R}}

\newcommand\bC{\mathbb{C}}

\newcommand\bT{\mathbb{T}}

\newcommand\sM{\mathcal{M}}

\newcommand\sB{\mathcal{B}}

\newcommand*\id{\mathop{}\!\mathrm{d}}

\newcommand{\tam}{\mathrm{argmin}}

\DeclareMathOperator*{\argmin}{arg\,min}

\newcommand{\tim}{\mathrm{Im}}

\newcommand{\ls}{\langle}
\newcommand{\rs}{\rangle}

\newtheorem{lemma}{Lemma}[section]

  \setlist[itemize]{leftmargin=*}

\begin{document}
 \title{Projected gradient descent for non-convex sparse spike estimation}
 \author {Yann Traonmilin$^{1,2}$, Jean-Fran\c cois Aujol$^{2}$ and Arthur Leclaire$^{2}$ \thanks{$^1$CNRS, $^2$Univ. Bordeaux, Bordeaux INP, CNRS,  IMB, UMR 5251, F-33400 Talence, France.}}

\maketitle
\begin{abstract}
We propose a new algorithm for sparse spike estimation from Fourier measurements. Based on theoretical results on non-convex optimization techniques for off-the-grid sparse spike estimation, we present a  projected gradient descent algorithm coupled with a back-projection initialization procedure. Our algorithm permits to estimate the positions of large numbers of Diracs in 2d from random Fourier measurements.  We present, along with the algorithm, theoretical qualitative insights explaining the success of our algorithm. This opens a new direction for practical off-the-grid spike estimation with theoretical guarantees in imaging applications.
\end{abstract}


\vspace*{-3mm}

\section{Introduction}

In the space $\sM=\sM(\bR^d)$ (respectively $\sM=\sM(\bT^d)$) of finite signed measures over $\bR^d$  (respectively the $d$-dimensional torus $\bT^d$), we aim  at recovering a superposition of impulsive sources $x_0 = \sum_{i=1}^k a_i \delta_{t_i} \in \sM$ from the measurements 
\begin{equation}
 y= Ax_0 + e,
\end{equation}
where $\delta_{t_i}$ is the Dirac measure at position $t_i$, the operator $A$ is a linear observation operator from $\sM$ to $\bC^m$ , $y \in \bC^m$ are the $m$ noisy measurements and $e$ is a finite energy observation noise. This inverse problem (called spike super-resolution \cite{Candes_2014,Bhaskar_2013,Tang_2013,Castro_2015,Duval_2015}) models many imaging problems found in geophysics, microscopy, astronomy or even (compressive) machine learning~\cite{Keriven_2017}. 
Under a separation assumption on the positions of the Diracs, i.e when $x_0$ is in a set $\Sigma_{k,\epsilon}$ of sums of $k$ $\epsilon$-separated Diracs with bounded support,  it has been shown that $x_0$ can be estimated by solving a non-convex problem  as long as $A$ is an appropriately designed measurement process. This ideal non-convex minimization is: 
\begin{equation} \label{eq:minimization}
    x^* \in \underset{x \in \Sigma_{k,\epsilon}}{\tam} \|Ax-y\|_2^2 .
\end{equation}
 Recovery guarantees for this problem are of the form 
\begin{equation}\label{eq:perf_bound}
 \|x^*-x_0\|_K\leq C \|e\|_2,
\end{equation}
where $\|\cdot\|_K$ is a kernel norm on $\sM$ that measures distances in $\sM$ at a given high resolution described by the kernel (in most of the literature $K$ is either a Fejér \cite{Candes_2014} or  Gaussian kernel and $\|\sum_i a_i \delta_{t_i}\|_K^2 = \sum_{i,j} a_ia_j K(t_i-t_j)$ \cite{Keriven_2018,Gribonval_2017}) . Mathematically, \eqref{eq:perf_bound} is guaranteed if the measurement operator $A$ has a restricted isometry property on $ \Sigma_{k,\epsilon}- \Sigma_{k,\epsilon}$ (the set of differences of elements of $\Sigma_{k,\epsilon}$)~\cite{Gribonval_2017}. This property is typically obtained when the number of measurements is sufficient. For example, recovery guarantees are obtained when $m \geq O(\frac{1}{\epsilon^d})$ for regular low frequency Fourier measurements on the torus~\cite{Candes_2014} and when $m \geq O(k^2d(\log(k))^2 \log(kd/\epsilon) )$ for random Fourier measurements on $\bR^d$~\cite{Gribonval_2017}. 

Recent advances in this field proposed a convex relaxation of the problem in the space of measures \cite{Candes_2014,Castro_2015}. While giving theoretical recovery guarantees, these methods are not convex with respect to the parameters due to a polynomial root finding step. Moreover, they rely on a SDP relaxation of a dual formulation, thus squaring the size of the problem. Sliding Frank-Wolfe/conditional gradient methods were also proposed but suffer from increased complexity when the number of spikes increases~\cite{Bredies_2013,Denoyelle_2019}. Methods based on structured low rank Toeplitz approximation are also difficult to extend to higher dimensions \cite{Condat_2015}.   Other methods based on greedy heuristics  (CL-OMP for compressive $k$-means~\cite{Keriven_2017}) have been proposed (very close in practice to the sliding Frank-Wolfe method) but they still lack theoretical justifications in this context even if some first theoretical results are emerging for some particular measurement methods~\cite{Elvira_2019}. 

In this paper, we propose a practical method to solve the non-convex minimization problem~\eqref{eq:minimization} for a large number of Diracs in imaging problems. Of course, at first sight, it is not possible to solve this  problem efficiently. However, we  justify qualitatively  why  our method  succeeds. This justification relies on  the separation assumption on $x_0$ and the assumption that there are enough  measurements of $x_0$. We also give numerical experiments validating the method. One of the main practical advantages of our method is its ability to perform off-the-grid spike estimation from random Fourier measurements with a good scaling with respect to the number of spikes. With this proof of concept, we can estimate many spikes in two dimensions from compressive measurements, yielding potential applications in fields such as astronomy or microscopy where the sum of spikes model is relevant.

Our method, following insights from the literature on non-convex optimization for low-dimensional models \cite{Waldspurger_2018,Ling_2017,Cambareri_2018,Chi_2019,Chizat_2019}, relies on two steps: 
\begin{itemize}
 \item  Overparametrized  initialization by hard-thresholded back-projection: we propose an initialization step that permits a good first estimation of the positions of the Diracs. 
 \item  Projected gradient descent algorithm in the parameter space: the idea of projected gradient descent for low-dimensional model recovery has shown its benefits in the finite dimensional case~\cite{Blumensath_2011,Golbabaee_2018}. We adapt this idea to  sparse spike recovery.  From \cite{Traonmilin_2018},  the global minimizer of~\eqref{eq:minimization} can be recovered by unconstrained gradient descent as long as the initialization lies in an explicit basin of attraction of the global minimizer. It was also shown that projecting on the separation constraint improves the control on the Hessian of the function we minimize. However, no practical way to perform a projection and no implementation were proposed. 
\end{itemize}

\noindent\textbf{Contributions.} After recalling  the context of non-convex sparse spike estimation, we propose a new practical projected gradient descent algorithm for sparse spike estimation. The simple gradient descent is already used as a refinement step in greedy algorithms. We show experimentally and justify qualitatively that adding a projection step and using an appropriate initialization leads to a global convergence. 
\begin{itemize}
 \item In Section~\ref{sec:algorithm}, we describe  our practical projected gradient descent algorithm and its implementation details; 
 \item In Section~\ref{sec:init}, we give  a grid based  initialization  using hard-thresholded back-projection. A qualitative analysis shows that when the number of measurements is large enough,  our initialization approximates well the Diracs positions;
 \item In Section~\ref{sec:exp}, we show  the practical  benefit of the projection in the descent algorithm and its application to the estimation of large number of Diracs in 2 dimensions.
\end{itemize}

\vspace*{-2mm}

\section{Theoretical background and algorithm description}\label{sec:algorithm} 
 
\subsection{Measurements and parameter space }
   The operator $A$ is a linear operator modeling $m$ measurements in $\bC^m$ ( $\tim A \subset \bC^m$ ) on the space of measures on a domain~$E$ (either $E = \bR^d$ or $E= \bT^d$) defined by: 
\begin{equation} \label{eq:distribution}
\forall l = 1, \ldots, m, \quad (Ax)_l =  \int_{E} \alpha_l(t) \id x(t) ,
\end{equation}
where $(\alpha_l)_{l=1}^m$ is a collection  of (weighted) Fourier measurements: $\alpha_l(t) =   c_l e^{-j \ls \omega_l,t\rs } $ for some chosen  frequencies $\omega_l \in \bR^d$ and frequency dependent weights $c_l \in \bR$ (the $c_l$ are mostly of theoretical interest for the study of recovery guarantees~\cite{Gribonval_2017} but can be set to $1$ in practice). The model set of $\epsilon$-separated Diracs with $\epsilon >0$ is:
\begin{equation}
\begin{split}
\Sigma_{k,\epsilon} := \left\{ \sum_{r=1}^k a_r \delta_{t_r} : \;  a \in \bR^k, t_r \in  \sB_2(R), \right. \\
\left. \forall  r \neq l, \|t_r-t_l\|_2 \geq \epsilon \right\},\\
\end{split}
\end{equation}
where  $\sB_2(R) = \{t \in \bR^d: \|t\|_2\leq R \}$ is the $\ell^2$ ball of radius $R$ centered in $0$ in~$\bR^d$. We consider the following parametrization of $\Sigma_{k,\epsilon}$:  for any $\theta= (a_{1},.., a_{k}, t_{1},..,t_{k}) \in \bR^{k(d+1)}$, we define $\phi(\theta) = \sum_{i=1}^k a_i \delta_{t_i}$, and we set
\begin{equation} 
  \Theta_{k,\epsilon}:= \phi^{-1}(\Sigma_{k,\epsilon}),
\end{equation}
the reciprocal image of $\Sigma_{k,\epsilon}$ by $\phi$. Note that any parametrization of elements of $\Sigma_{k,\epsilon}$ is invariant by permutation of the positions. This is not a problem in practice for the convergence of descent algorithms. 
We define the parametrized functional 
\begin{equation}
  g(\theta)  :=  \|A\phi(\theta)-y\|_2^2  
\end{equation}

and consider the problem
\begin{equation} \label{eq:minimization2}
    \theta^* \in \argmin_{ \theta \in \Theta_{k,\epsilon}}  g(\theta) .
\end{equation}

Since the $\alpha_l$ are smooth, $g$ is a smooth function. Note that performing the minimization~\eqref{eq:minimization2} allows to recover the minima of the ideal minimization~\eqref{eq:minimization}, yielding stable recovery guarantees under a restricted isometry assumption on $A$ which is verified when $m \geq O(k^2d(\log(k))^2 \log(kd/\epsilon))$ for adequately chosen Gaussian random Fourier measurements (on $\bR^d$) and $m \geq O(\frac{1}{\epsilon^d})$ for regular Fourier measurements on $\bT^d$.  
In \cite{Traonmilin_2018}, it has been shown that the simple gradient descent converges  (without projection) to the global minimum of $g$ as long as the initialization falls into an explicit basin of attraction of this global minimum. It was also shown that the projection on the separation constraint  improves the control on the Hessian on $g$ and subsequently the convergence of the descent algorithm.

\subsection{Projected gradient descent in the parameter space }

For a user-defined initial number of Diracs $k_{in}$, we consider the following iterations: 
\begin{equation}
 \begin{split}
  \theta_{n+1} &= P_{\Theta_{k_{in},\epsilon}}(\theta_n - \tau_n \nabla g(\theta_n))
 \end{split}
\end{equation}
where $ P_{\Theta_{k_{in},\epsilon}}$ is a projection on the separation constraint,
 (notice that there may be several solutions in  $\Theta_{k_{in},\epsilon}$) and $\tau_n$ is the step size at iteration $n$. The projection  $P_{\Theta_{k_{in},\epsilon}}(\theta)$ could be defined naturally as a solution of the minimization problem 
$\inf_{\tilde \theta \in 
 \Theta_{k_{in},\epsilon}} \|\phi(\tilde \theta) - \phi(\theta)   \|_K$. Unfortunately this optimization is not convex.  We  propose instead a heuristic (see Algorithm~\ref{alg:Projection}) for $ P_{\Theta_{k_{in},\epsilon}}$ that consists in merging Diracs that are not $\epsilon$-separated.  

\begin{algorithm}

\DontPrintSemicolon
\KwIn{ List $\Theta=  (a_i, t_i)_i $ of amplitudes and positions ordered by decreasing absolute amplitudes} 
  \For{ $i \geq 1$}{
  \For{ $j >i$}{
    \If{$\|t_i-t_j\| < \epsilon$}{ 
     
     $a_i = a_i+a_j$;
     
     $t_i = \frac{|a_i|t_i + |a_j|t_j}{|a_i|+|a_j|}$
     
     Remove $(a_j,t_j)$ from $\Theta$ 
    }
}
   }
   \KwOut{ List $\Theta=  (a_i, t_i)_i $ of amplitudes and positions of projected  spikes}
   \vspace{2mm}
\caption{Heuristic for the projection  $P_{\Theta_{k_{in},\epsilon}}$}
\label{alg:Projection}
\end{algorithm}

Since we take the barycenter of the positions,  if  a set of Diracs that are at a distance at most $\epsilon$ of a true position in $x_0$ is merged, the merged result will be within this distance. We use the ordering by decreasing amplitude to avoid that a low amplitude spike pulls a meaningful high-amplitude spike away from a true position. After this projection step, we pursue the descent with the remaining number of Diracs. Note that we overparametrize with $k_{in}$ the number of Diracs in the descent to ensure the recovery of all positions in $x_0$ (see also the next section). In practice, we implement the projected gradient descent as follows. 
\begin{itemize}
 \item As suggested in \cite{Traonmilin_2018},  to avoid balancing problems between amplitudes and positions, we alternate descent steps between amplitudes and positions.
 \item To find the step size $\tau_n$, we perform a line search to minimize the value of the function $g$.
 \item We start to project after a few iterations (20 iterations in our experiments) of the gradient descent so that spikes have already started clustering together towards the solution.
\end{itemize}

From \cite{Traonmilin_2018}, this algorithm will converge as soon as the initialization falls into a basin of attraction of  global minimum of $g$. The basins of attraction get larger as the number of measurements increases (up to a fundamental limit depending on the separation $\epsilon$ and the amplitudes in $x_0$). 
\vspace*{-2mm}

\section{Overparametrized  initialization by hard-thresholded back-projection} \label{sec:init}

The idea of using back-projection of measurements was used for non-convex optimization in the context of phase recovery~\cite{Waldspurger_2018} and blind deconvolution~\cite{Cambareri_2018} with so-called spectral initialization techniques, where a leading eigenvector of a matrix constructed with back-projections is used as initialization.  As we measure the signal $x_0$ at some frequencies $\omega_l$, a way to recover an estimation of the signal is to back-project the whole irregular spectrum on a grid $\Gamma$ that samples $\sB_2(R)$ at a given precision $\epsilon_g$ (to be chosen later). 
For Fourier measurements $y$ at frequencies $(\omega_l)_{l=1,m}$, we calculate $z_\Gamma = B_\Gamma y$ where $B_\Gamma$ is the linear operator back-projecting the Fourier measurements on a  grid in the spatial domain $\bR^d$:
\begin{equation}
z_\Gamma= B_\Gamma y := \sum_{s_i \in \Gamma} z_{\Gamma,i} \delta_{s_i}
\end{equation}
where the $s_i \in \Gamma$ are the grid positions and
\begin{equation}
z_{\Gamma,i}=  \sum_{l}y_l d_l e^{j \ls \omega_l,s_i\rs}
\end{equation}
for some appropriate weights $d_l$ to be chosen in the next section. We can show in the noiseless case with Lemmas~\ref{lem:loc_fejer} and ~\ref{lem:loc_gaussian} that when the number of measurements increases and the grid for initialization gets finer, the original positions of Diracs get better approximated.  All ``on-the-grid" methods such as least-squares estimation or the LASSO are ways to back-project measurements. As we aim at a fast algorithm, and since the energy of Diracs is well localized by the initialization we then perform overparametrized hard thresholding  of the back-projection. We propose the initialization $\theta_{init}$ defined by
\begin{equation}\label{eq:initialization}
 \phi(\theta_{init}):=x_{init} = H_{k_{in} }(B_\Gamma y)
\end{equation}
where for  $|z_{\Gamma,j_1}| \geq |z_{\Gamma,j_2}| \geq .... |z_{\Gamma,j_n}|$, we have $H_{k_{in}}(z_\Gamma) = \sum_{i=1}^{k_{in}} z_{\Gamma,j_i} \delta_{s_{j_i}}$. \\

\noindent\textbf{Ideal back-projection and sampling:} In the context of Diracs recovery our initialization by hard-thresholded back-projection is a sampling of an ideal back-projection. Let $B$ the operator from $\bC^m$ to $\sM$ defined for  $z = By$  by
\begin{equation} \label{d_l}
 z(t) = \sum_{l=1}^m d_l y_l e^{j \ls \omega_l, t \rs} 
\end{equation}
We call $z$ an ideal back-projection because $z_\Gamma = S_{\Gamma}z$ where $S_\Gamma$ is the sampling on the grid $\Gamma$: for a measure $x$ with a continuous density $\chi$ (i.e. $\id x (t) = \chi(t)  \id t $), we define the sampling operation $S_{\Gamma}(x) = \sum_{t_i \in \Gamma} \chi(t_i)\delta_{t_i}$. Also, the measure $z = By$ has a smooth  density as it is a finite sum of complex exponentials.

We first show (proofs are in the supplementary material) that for the right choice of weights $d_l$, the energy of $z = BAx$ (where $x = \sum_{i=1}^k a_i \delta_{t_i}$) is localized around the positions $t_i$ in both the regular Fourier sampling  on the torus case, and the random Fourier sampling on $\bR^d$ case. We show the following results for the Fejér and Gaussian kernels as they are typically used in the literature for deterministic \cite{Candes_2014} and random \cite{Gribonval_2017} Fourier sampling. 

\begin{figure}[ht!]
 \vspace*{-2mm}
 \begin{center}
  \includegraphics[width=0.35\linewidth]{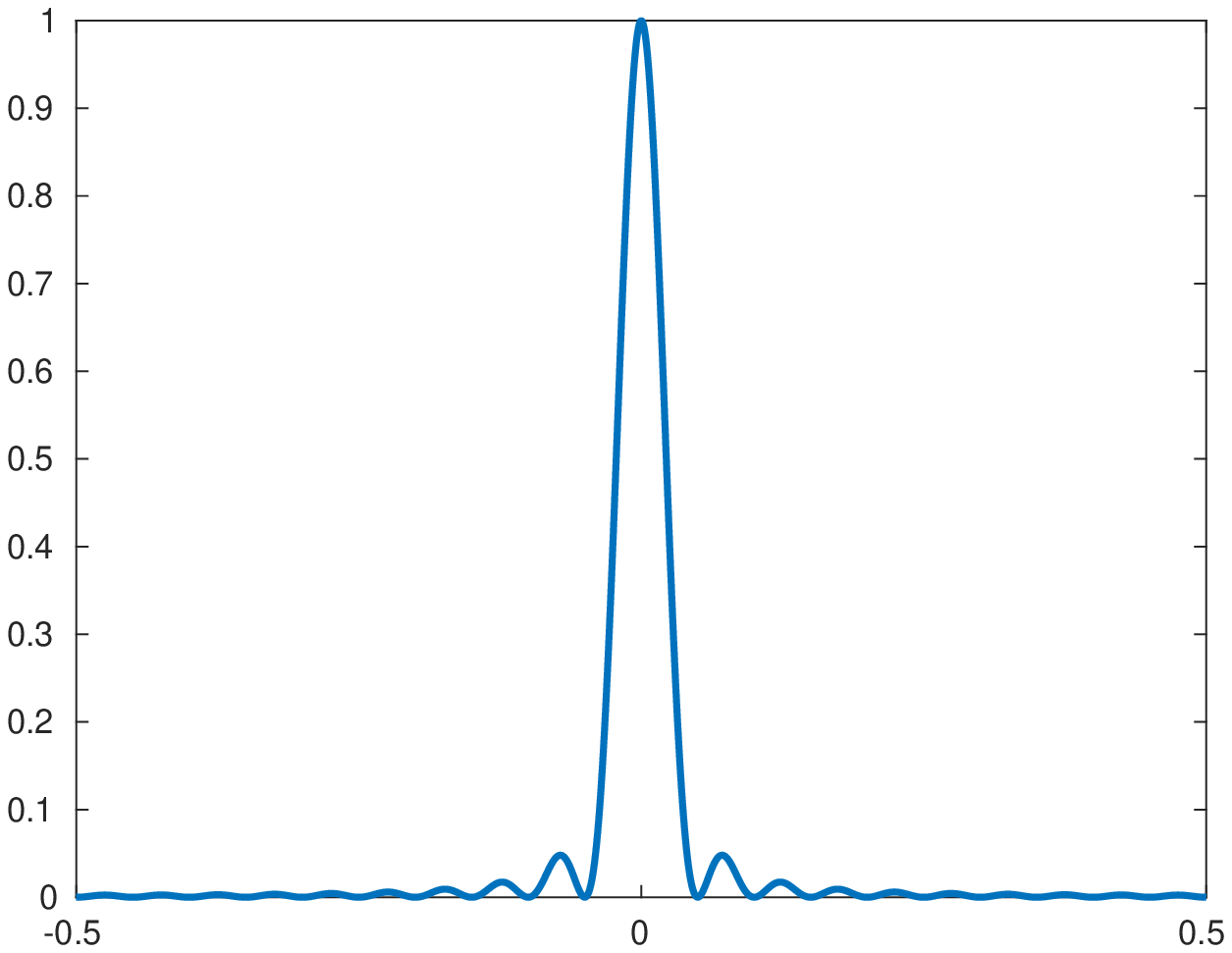}
  \includegraphics[width=0.35\linewidth]{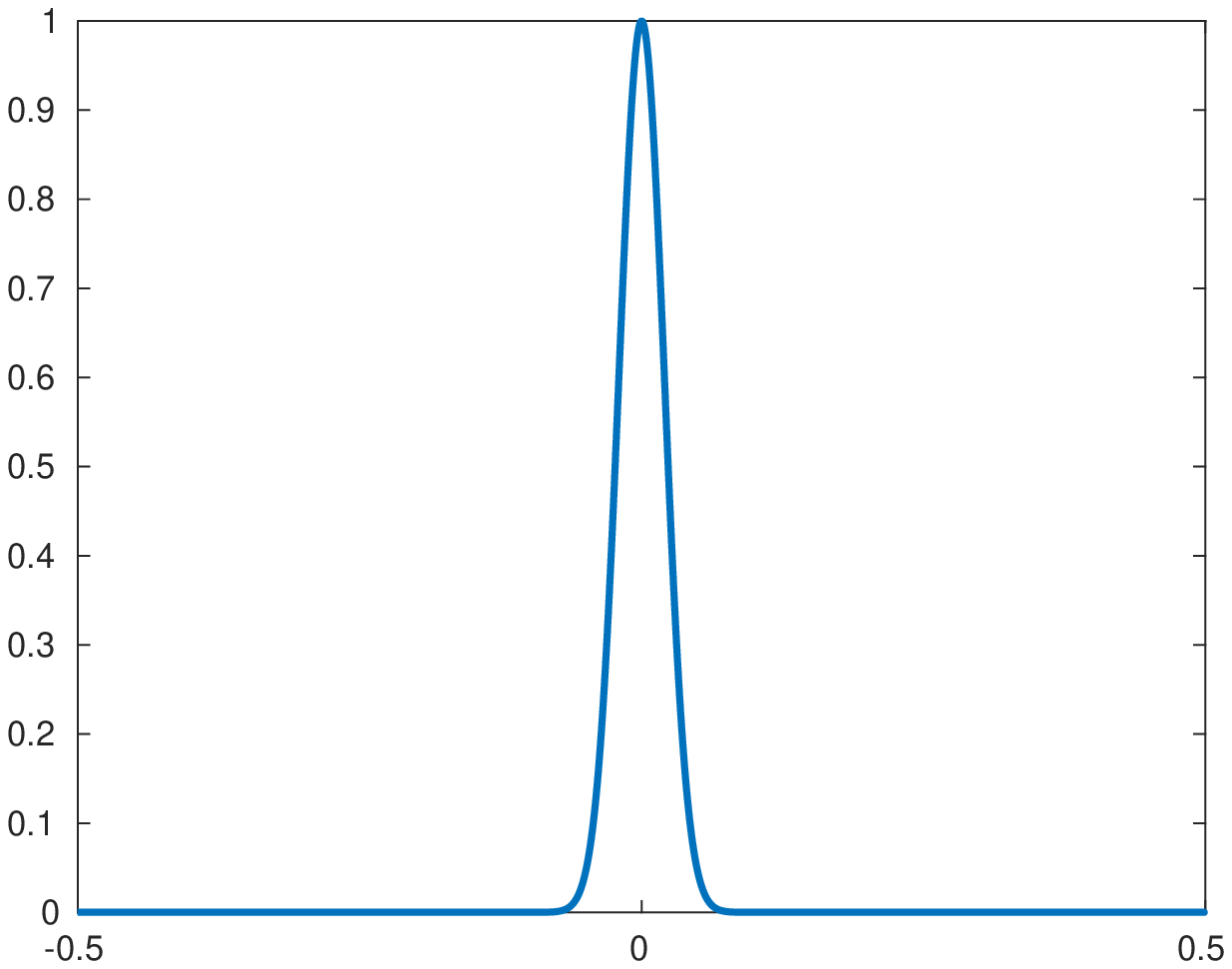}
 \end{center}
 \vspace*{-5mm}
\caption{Kernels used for a separation $\epsilon = 0.1$. Left: On the torus the Fejér kernel of maximum frequency $\frac{2}{\epsilon}$. Right: on $\bR$, the Gaussian kernel of parameter $\sigma =\frac{1}{500\epsilon}$ }\label{fig:kernels}\vspace*{-5mm}
\end{figure}

\begin{lemma}\label{lem:loc_fejer}
On $\sM(\bT^d)$,  we choose $A$ such that $(\omega_l)_{l=1,m}$ is a regular sampling of $[-\omega_{max},\omega_{max}]^d$ with  $\omega_l \in 2\pi.\sZ^d $.  In~\eqref{d_l}, take $d_l =\hat{K_f}(\omega_l)/((2 \pi)^dc_l)$ where $\hat{K_f}$ is the Fourier transform of the Fejér kernel $K_f$ on the torus whose Fourier spectrum support is $(\omega_l)_{l=1,m}$, then
\begin{equation}
\begin{split}
z(t) &=  \sum_{i=1}^k a_i K_f(t-t_i).\\
\end{split}
\end{equation}
\end{lemma}
This immediate lemma states that on the torus, measuring low frequencies is equivalent to measuring a low-pass filtered signal in the time domain. For example, the low-pass Fejér filter is shown Figure~\ref{fig:kernels}. Also, this result holds for any kernel with spectrum supported on the $\omega_l$.  Sampling  frequencies regularly with maximum frequency $\omega_{max} \geq O(\frac{1}{\epsilon^d})$ guarantees recovery with convex relaxation methods. For random Fourier sampling, we  look at the expected value of $z$  and control its variance with respect to the distribution of the $\omega_l$.

\begin{lemma}\label{lem:loc_gaussian}
On $\sM(\bR^d)$,  we choose $A$  such that  the $\omega_l$ are $m$ i.i.d  random variables with a Gaussian distribution with density $G(\omega_r)  = \frac{\sigma^d}{(\sqrt{2\pi})^d} e^{-  \frac{\sigma^2}{2}\|\omega_r\|_2^2 }$.  Let $K_g(t) = e^{-  \frac{\|t\|_2^2}{2\sigma^2} } $. In~\eqref{d_l}, take $d_l = 1 /( m c_l)$  then 
\begin{equation}
\begin{split}
E(z(t)) &=  \sum_{i=1}^k a_i K_g(t-t_i)\\
\end{split}
\end{equation}
\begin{equation}
\begin{split}
E( |z(t) -E(z(t))|^2) &= -\frac{1}{m} |E(z(t))|^2   + \frac{1}{m} \|x_0\|_{K_g}^2 \\
\end{split}
\end{equation}
where $\|x_0\|_{K_g}^2$ is the norm associated with the kernel $K_g$.
\end{lemma}
 
Similarly to the regular sampling, the energy of the  expected value of $z$ is concentrated around the positions $t_i$ (see Figure~\ref{fig:kernels}). In \cite{Gribonval_2017} the frequency distribution scales as the inverse of the kernel precision, i.e. the kernel parameter $\sigma$ of the kernel $h(t) =e^{-\|t\|_2^2/(2 \sigma^2)}$ is chosen as $O(1/\epsilon)$.  The control of the variance shows that when the number of measurements increases, the back-projection of these measurements to the space of measures are closer to the ideal initialization which is the expected value of $z$.  In practice we set the number of measurements using a rule $m = \mu kd $ with a user defined multiplicative parameter $\mu$ that does not depend on the dimension of the problem. The quality of the initialization is thus directly linked to  $\mu$. Finally the following lemma  makes sure that as the grid gets finer we recover all the energy of the ideal back-projection that lies within the domain sampled by the grid. 
\begin{lemma}
 Let $z_d = (z_{\Gamma,i})_{t_i \in \Gamma}$ where $\Gamma$ is a grid with step size $\epsilon_g$. Then  $\|\sqrt{\epsilon_g^d}z_d\|_2^2 \to_{\epsilon_g \to 0} \|z\|_{L^2(\sB_2(R))}^2 $.
\end{lemma}

We considered the noiseless case.  The noisy case just adds a noise term with energy controlled by the noise energy level $\|e\|_2$ because $B_\Gamma$ is a Fourier back-projection.
\vspace*{-3mm}

\section{Numerical Experiments}  \label{sec:exp}

We first run the algorithm on few Diracs in 2d to illustrate the added benefit of the projection.  We  then show results with many Diracs in 2d to show the  computational feasibility  of projected gradient descent for imaging applications. We perform the experiments in the noiseless case with a stopping criterion based on the value of function $g$ and leave the study of the impact of the noise for future work. The Matlab code used to generate these experiments is available at \cite{Traonmilin_2019code}. \\
   \vspace*{-2mm}

\begin{figure}
 \begin{center}
  \includegraphics[width=0.45\linewidth]{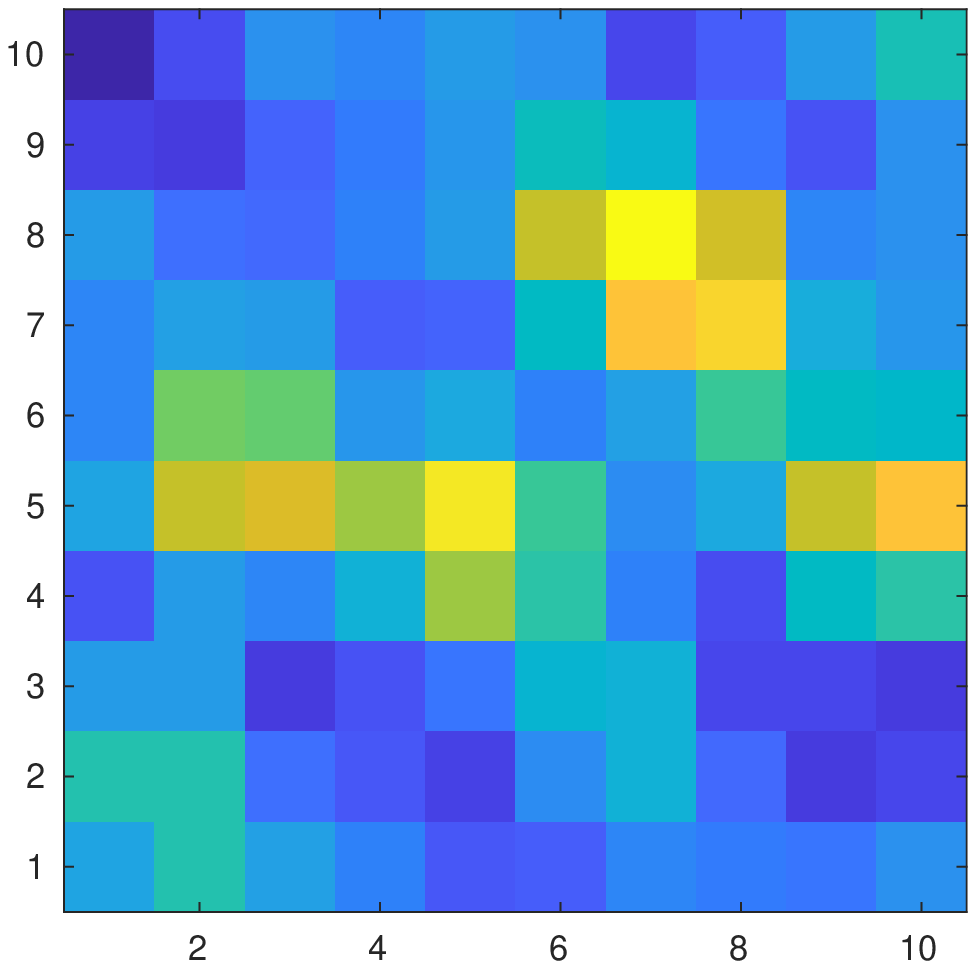} \includegraphics[width=0.45\linewidth]{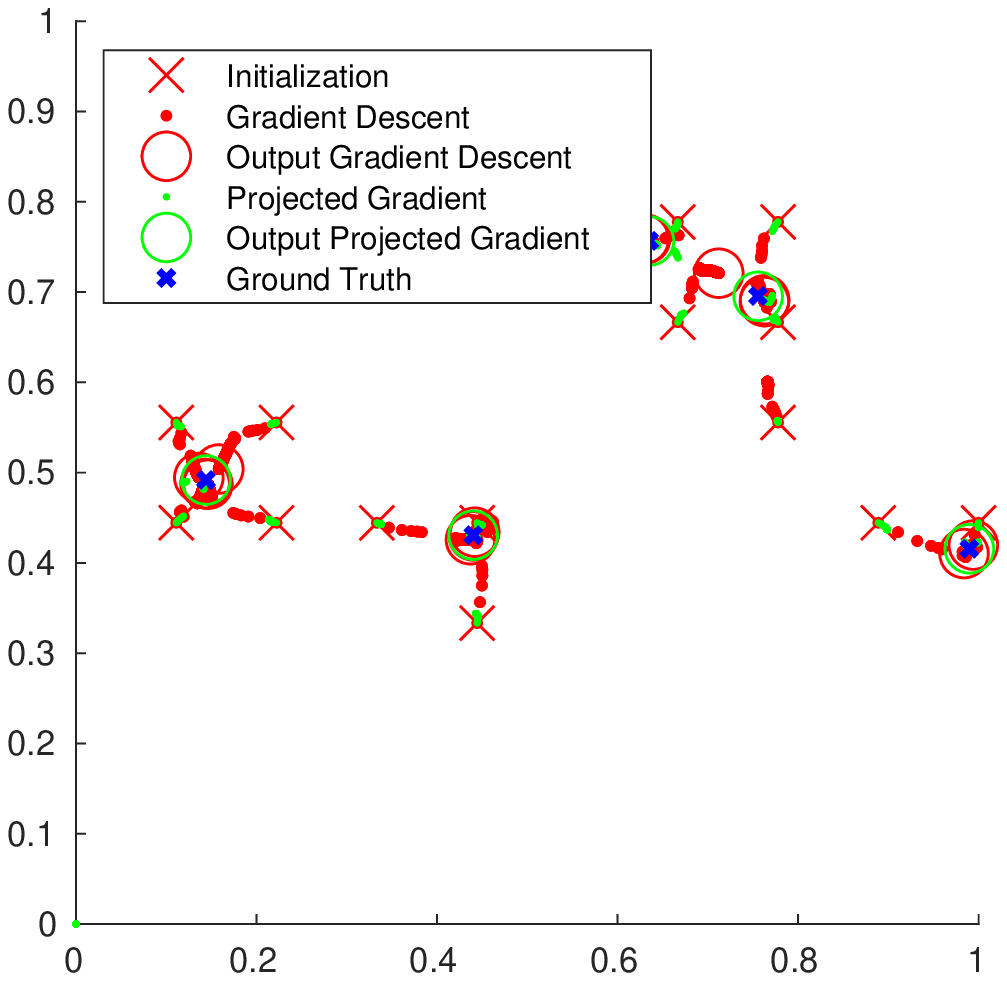}
 \end{center}
  \vspace*{-5mm}
\caption{Result for a few spikes in 2d. Left: back-projection of measurements on a grid. Right: Initialization, gradient descent and projected gradient descent trajectories.}\label{fig:proj_grad_simple}\vspace*{-5mm}
\end{figure}

\noindent\textbf{Illustration with few Diracs:}  
As a first proof of concept we run the algorithm with the recovery of $5$ Diracs in 2 dimensions from $m=120$ Gaussian random measurements. The trajectories of $500$ iterations of the gradient descent and projected gradient descent are represented in Figure~\ref{fig:proj_grad_simple}. 
We observe that while the gradient descent with overparametrized initialization might converge with a large number of iterations, the projection step greatly accelerates the convergence. \\
 
   \vspace*{-2mm}

\begin{figure}[!ht]
   \vspace*{-3mm}
 \begin{center}
   \includegraphics[width=0.45\linewidth]{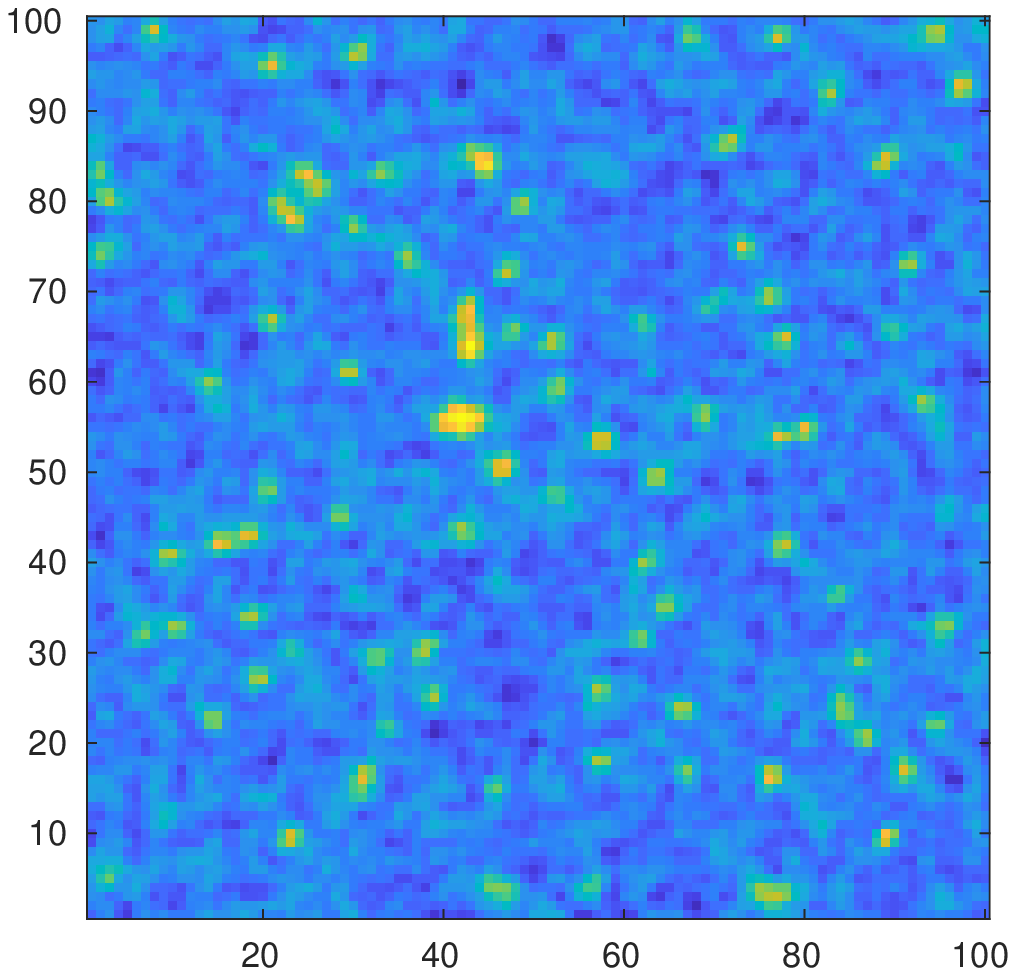}
  \includegraphics[width=0.45\linewidth]{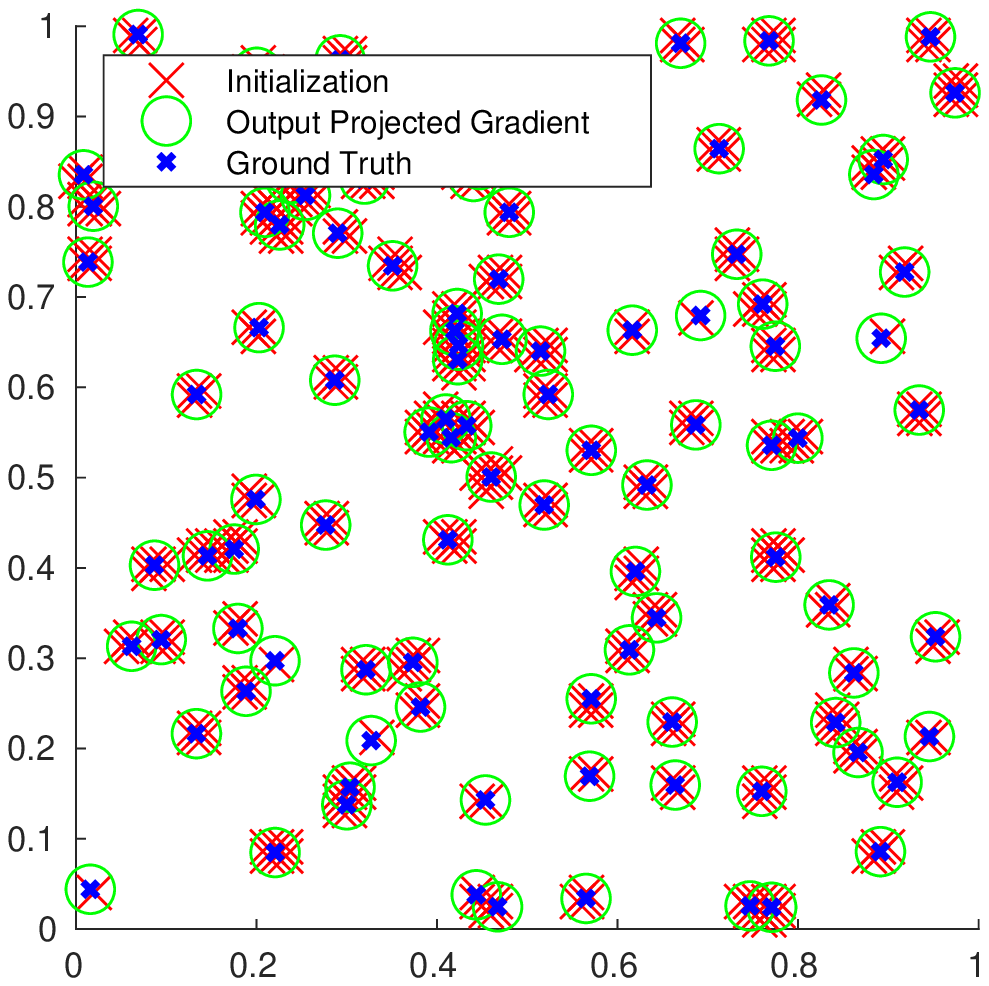}
 \end{center}
 \vspace*{-5mm}
\caption{Result for 100 spikes in 2d.  Left: back-projection of measurements on a grid. Right: Initialization, and projected gradient descent trajectories.}\label{fig:proj_grad_big}
\vspace*{-5mm}
\end{figure}

\noindent\textbf{Estimation of 100 Diracs in 2d:} 
We recover 100 Diracs, with a separation $0.01$ on the square $[0, 1] \times [0, 1]$ from $ m = 2000$ compressive Gaussian measurements (we would need $\approx 10000$ regular measurements to obtain a separation $0.01$). In  practice, the grid $\Gamma$ must be fine enough to overparametrize the number of Diracs with a good sampling of the ideal back-projection. If $\epsilon_g$ is too small, the number of initial Diracs needed to sample the energy gets larger, leading to an increased cost in the first iterations   of the gradient descent. In this example we use $\epsilon_g = \epsilon$ and use $k_{in} = 4k$. We observe in Figure~\ref{fig:proj_grad_big} that with these parameters all the Diracs positions are well estimated after $184$ iterations (convergence criterion met) of our algorithm. Similarly to our first example, we  observe that spikes that are not separated in the back-projection  on the grid are well estimated by our algorithm.

\subsection{Complexity} 
The cost of our algorithm is the sum of the cost of the initialization and the cost of the projected gradient descent. The back-projection on the grid scales as $O((1/\epsilon_g)^d)$ (irregular Fourier transform on a grid), but it is done only once and fast transform techniques could be investigated. With our strategy, this cost seems unavoidable as we want to localize Diracs off-the-grid with a separation $\epsilon$ (doing the same on the grid would have this exponential scaling with respect to the dimension and the separation). Our algorithm stays tractable when the  dimension $d$ of the domain of the positions of the Diracs is not too large, which is the case in 2d or 3d imaging applications. This cost $O((1/\epsilon_g)^d)$ would also be impossible to avoid in an eventual vizualisation of the full recovered image over a precise grid on $\bR^d$. Note also that in practice our proposed initialization could be replaced by any statisfying overparametrized initialization, i.e. any state-of-the-art on-the-grid estimation technique could benefit from an added projected gradient descent. 

The cost of the projected gradient descent is  $O(n_{it}C_{\nabla})$ where $C_\nabla$ is the cost the calculation of the gradient. This cost is  of the order of the calculation of the  $m$ Fourier measurements for the current number of Diracs in the descent (close to  $k$ after a few iterations). For the experiment with 100 spikes, ouralgorithm successfully completed the estimation of the spikes in 5.9 minutes (Matlab implementation) while the CL-OMP algorithm proposed by \cite{Keriven_2016} (their Matlab implemenation) took 30.5 minutes to complete the successfull estimation of all the spikes on a laptop for office purpose. 
\vspace*{-2mm}

\section{Conclusion} 
We gave a practical algorithm to perform off-the-grid sparse spike estimation. This proof-of-concept shows that it is possible to estimate efficiently a large number of Diracs in imaging applications with some strong theoretical insights of success guarantees. Future research directions are: 
\begin{itemize}
 \item Full theoretical convergence proof of the algorithm with sufficient conditions on the number of measurements. The main question is to know if it is possible to have a convergence guarantee without the computational cost $O((1/\epsilon_g)^d)$
 \item Investigate other methods for reducing the number of parameters after the back-projection on a grid and accelarate the descent (quasi-Newton schemes ).
 \item Study the algorithm stability to noise and modeling error with respect to the number of measurements. 
\end{itemize}

\bibliographystyle{abbrv}
 \bibliography{non_convex_SR_SPL}
\end{document}



\maketitle
 \onecolumn

\section{Definitions and properties}

For $x_0 = \sum_{i=1,k} a_i \delta_{t_i}$, we have $y_l=(Ax_0)_l = \sum_{i=1,k} a_i e^{-j \ls w_l , t_i\rs }$.

\noindent The Fourier transform of a function $f$ supported on $\bR^d$  is 

\begin{equation}
 \hat{f}(\omega) = \int_{t \in bR^d}  e^{-j \ls \omega, t \rs } f(t)dt
\end{equation}

\noindent Let  $E =\bR^d$. The Gaussian function $G$ defined by $G(u) = \frac{\sigma^d}{(\sqrt{2 \pi})^d} e^{- \frac{\sigma^2}{2} \|u\|_2^2}$. Then  $\hat{G}(v) = e^{- \frac{1}{2\sigma^2} \|v||_2^2} $.

\section{Proofs}
\begin{proof}[Proof of Lemma III.1.]
Using the fact that the $\omega_l$ sample exactly the support of the F\'ejer Kernel and that, by definition,
\begin{equation}
 K_f(t) = \sum_{l=1,m} \hat{K}(\omega_l) e^{j \ls \omega_l, t \rs },
 \end{equation}
we have 
\begin{equation}
\begin{split}
z(t) &= \sum_{l=1,m} d_l (\sum_{i=1,k} a_i  c_l  e^{-j \ls \omega_l,t_i \rs})   e^{j \ls \omega_l,t \rs}   \\
 &=  \sum_{i=1,k} a_i \sum_{l=1,m} d_l c_l    e^{j \ls \omega_l,t-t_i \rs}   \\
 &=  \sum_{i=1,k} a_i K_f(t-t_i).\\
\end{split}
\end{equation}

\end{proof}

 \begin{proof}[Proof of Lemma III.2.]
 Using the properties of the Fourier transform $\hat{G}$ of the Gaussian function $G$ and $\bE$ being the expectation with respect to the distribution of the frequencies $\omega_l$, we have 
  \begin{equation}
\begin{split}
\bE(z(t)) &=  \sum_{i=1,k} a_i \sum_{l=1,m} d_l c_l  \bE (e^{j \ls \omega_l,t-t_i \rs} )\\
 &=   \sum_{i=1,k} a_i \sum_{l=1,m} d_l c_l \hat{G}(t_i-t)\\
 &=   \sum_{i=1,k} a_i   K_g(t_i-t).
\end{split}
\end{equation}

 Using the formula $\bE( |z(t) -E(z(t))|^2) =  \bE( |z(t)|^2) - |E(z(t))|^2  $ and the previous Lemma, we just need to calculate 

  \begin{equation}
\begin{split}
\bE(|z(t)|^2) &=  \bE \left( \left|\sum_{i=1,k} a_i \sum_{l=1,m} d_l c_l  e^{j \ls \omega_l,t-t_i \rs} \right|^2 \right)\\ 
&=  \bE \left( \sum_{i_1=1,k} a_{i_1} \sum_{l=1,m} d_l c_l  e^{j \ls \omega_l,t-t_{i_1} \rs} \sum_{i_2=1,k} a_{i_2} \sum_{l=1,m} \bar{d}_l \bar{c}_l  e^{-j \ls \omega_l,t-t_{i_2} \rs} \right) \\
&=   \sum_{i_1=1,k}\sum_{i_2=1,k}   a_{i_1} a_{i_2} \sum_{l=1,m} \sum_{r=1,m} d_l c_l   \bar{d}_r \bar{c}_r  \bE \left( e^{j \ls \omega_l,t-t_{i_1} \rs}   e^{-j \ls \omega_r,t-t_{i_2} \rs} \right)\\
\end{split}
\end{equation}
For  $r = l$, 
  \begin{equation}
\begin{split}
\bE  \left(e^{j \ls \omega_l,t-t_{i_1} \rs}   e^{-j \ls \omega_r,t-t_{i_2} \rs}\right) &= \hat{G}(t_{i_2}-t_{i_1})\\
\end{split}
\end{equation}
For  $r \neq l$, 
  \begin{equation}
\begin{split}
\bE  \left(e^{j \ls \omega_l,t-t_{i_1} \rs}   e^{-j \ls \omega_r,t-t_{i_2} \rs}\right) &=\bE  \left(e^{j \ls \omega_l,t-t_{i_1} \rs}  \right)  \bE  \left(e^{-j \ls \omega_r,t-t_{i_2} \rs}\right)  \\
 &= \hat{G}(t-t_{i_1})  \hat{G}(t-t_{i_2}) \\
\end{split}
\end{equation} 
Hence 
  \begin{equation}
\begin{split}
\bE(|z(t)|^2) =&   \sum_{i_1=1,k}\sum_{i_2=1,k}   a_{i_1} a_{i_2}  \left( \sum_{l=1,m} |d_l c_l|^2   \hat{G}(t_{i_2}-t_{i_1})  + \sum_{l\neq r}  d_l c_l   \bar{d}_r \bar{c}_r \hat{G}(t-t_{i_1})  \hat{G}(t-t_{i_2}) \right)\\
 =&   \sum_{i_1=1,k}\sum_{i_2=1,k}   a_{i_1} a_{i_2}  \left( \sum_{l=1,m} \frac{1}{m^2}K_g(t_{i_2}-t_{i_1})  + \sum_{l\neq r}  \frac{1}{m^2}   K_g(t-t_{i_1})  K_g(t-t_{i_2}) \right)\\
=&   \left(1-  \frac{1}{m}\right)  (\sum_{i=1,k} a_{i} K_g(t-t_i))^2  +  \frac{1}{m} \sum_{i_1=1,k}\sum_{i_2=1,k}   a_{i_1} a_{i_2}  \frac{1}{\ m}  K_g(t_{i_2}-t_{i_1}) \\
=& \left(1-  \frac{1}{m}\right)  |E(z(t))|^2   + \frac{1}{m} \|x_0\|_K^2 \\
\end{split}
\end{equation}
Going back to the variance, 

\begin{equation}
\begin{split}
\bE( |z(t) -E(z(t))|^2) &=  \frac{1}{m} |E(z(t))|^2   + \frac{1 }{m} \|x_0\|_{K_g}^2\\
\end{split}
\end{equation}

\end{proof}
\begin{proof}[Proof of Lemma III.3.]
\begin{equation}
 \|\sqrt{\epsilon_g^d}z_d\|_2^2 =  \epsilon_g^d\sum_{i\in\Gamma} |z(t_i)|^2
\end{equation}

As $z$ is continuous on $\sB_2(R)$, it is integrable over $\sB_2(R)$ and this sum is approximating the Riemann integral of $t \to |z(t)|^2$ over $\sB_2(R)$ as the grid step size $\epsilon_g$ of $\Gamma$ converges to 0. 
\end{proof}